\documentclass{JHEP3}

\usepackage{
cite,
amssymb,amsmath,epsf,graphics
}
\usepackage[dvips]{graphicx}

\numberwithin{equation}{section}
\newcommand{\al}{\alpha}

\newcommand{\ga}{\gamma}

\newcommand{\si}{\sigma}
\newcommand{\eps}{\epsilon}
\newcommand{\om}{\omega}

\newcommand{\vth}{\vartheta}
\newcommand{\G}{\Gamma}

\newcommand{\Si}{\Sigma}

\newcommand{\Om}{\Omega}

\newcommand{\cH}{{\cal H}}
\newcommand{\cI}{{\cal I}}

\newcommand{\cN}{{\cal N}}

\newcommand{\non}{\nonumber}
\newcommand{\Bz}{{\mathbb Z}}
\newcommand{\Br}{{\mathbb R}}
\newcommand{\Bc}{{\mathbb C}}

\newcommand{\ident}{1 \hspace{-1mm} {\rm l}}

\newcommand{\bq}{\bar{q}}

\newcommand{\bL}{\bar{L}}

\newcommand{\tf}{\tilde{f}}

\newcommand{\tpsi}{\tilde{\psi}}

\newcommand{\im}{{\rm Im}}
\newcommand{\re}{{\rm Re}}
\newcommand{\Tr}{{\rm Tr}}

\newcommand{\CY}{{Calabi-Yau~}}

\newcommand{\ie}{{\it i.e.}}

\title{Generalised discrete torsion and\\ mirror symmetry for
G$_{\mathbf 2}$ manifolds}

\renewcommand{\thefootnote}{\fnsymbol{footnote}}

\author{Matthias R.\ Gaberdiel\footnote{email: gaberdiel@itp.phys.ethz.ch} 
~and~ Peter Kaste\footnote{email: kaste@itp.phys.ethz.ch}\\
Institute for Theoretical Physics\\
ETH H\"onggerberg\\
CH--8093 Z\"urich, Switzerland}

\preprint{\hepth{0401125}}

\abstract{
A generalisation of discrete torsion is introduced in which
different discrete torsion phases are considered for the 
different fixed points or twist fields of a twisted sector. The
constraints that arise from modular invariance are analysed carefully.  
As an application we show how all the different resolutions of
the $T^7/\Bz_2^3$ orbifold of Joyce have an interpretation 
in terms of such generalised discrete torsion orbifolds. Furthermore,  
we show that these manifolds are pairwise identified under G$_2$
mirror symmetry. From a conformal field theory point of view, this
mirror symmetry arises from an automorphism of
the extended chiral algebra of the G$_2$ compactification.
}


\begin{document} 

\maketitle 

\renewcommand{\thefootnote}{\arabic{footnote}}
\setcounter{footnote}{0}

\section{Introduction}

It has been known for a long time that orbifolds in string theory are
not uniquely characterised in terms of the action of the orbifold
group on the states of the original theory. In fact, in
order to construct the theory one also has to specify the action of
the orbifold group $\Gamma$ in the various twisted sectors of the
theory, and in general this is not unambiguously defined. In
particular, as was pointed out by Vafa \cite{Vafa}, there is at least
the freedom to modify the action of $g$ in the $h$-twisted sector by a
phase $\epsilon(g,h)$. Provided that these phases correspond to a
2-cocycle $H^2(\Gamma,U(1))$, the resulting theory is modular
invariant and thus consistent (given that the original orbifold theory
was so). Thus if $H^2(\Gamma,U(1))\ne \{e\}$, the orbifold
construction is ambiguous, and needs to be specified further.

One may wonder whether the above ambiguity is the only ambiguity that
is consistent with modular invariance, or whether there are additional
possibilities in general. Clearly, any such additional possibilities
will depend on the specifics of the theory in question, and it will
therefore not be possible to give a general analysis as for the case
of conventional discrete torsion. However, it is nevertheless
interesting to understand whether there are such additional theories
in specific instances. 

Our interest in this problem arose from the analysis of orbifolds
that describe the compactification of IIA or IIB string theory on the 
G$_2$ manifolds of Joyce \cite{Joyce:1996a,Joyce:1996b}, in particular the 
family of nine G$_2$ manifolds obtained by inequivalent resolutions of
the $T^7/\Bz_2^3$ orbifold of \cite[chapter 12.3]{Joyce:book}.
It was shown
in \cite{Shatashvili:1994zw} how one of the nine possible resolutions
can be constructed in string theory. By
switching on (conventional) discrete torsion, another resolution was
found in \cite{Acharya1,Acharya2}, but it was not clear how to obtain
the remaining seven 
in terms of string theory. In this paper we
want to explain how these remaining resolutions can be
constructed. This involves a generalisation of the usual discrete
torsion construction in which different discrete torsion phases are
switched on for the different fixed points of a given twisted
sector. The constraint that the resulting theory must still be modular
invariant imposes some constraints on the choice of these phases, and 
we shall analyse them in detail. In fact, we shall find that there are
(up to some relabelling) precisely nine different string theories that
are allowed by these constraints, and that they correspond precisely
to the nine different resolutions found by Joyce.  
\smallskip

For a closely related Calabi-Yau manifold, it was shown in
\cite{Vafa:1995rv} that the orbifold with discrete torsion is related
to the same orbifold without discrete torsion by mirror symmetry
\cite{Lerche:1989uy}. In fact, mirror symmetry simply corresponds to
T-duality along three circles in this case. This suggests that
something similar may be true for the G$_2$ manifolds in
question. (The idea that some version of mirror symmetry should also
apply to G$_2$ manifolds was first proposed in
\cite{Joyce:1996b} and argued for on physical grounds in
\cite{Shatashvili:1994zw}.) 
It was shown in \cite{Acharya1,Acharya2}
that the theory with and without (conventional) discrete torsion are
indeed related by three T-dualities to one another, and this suggests
that one should regard them as mirror partners. 

For Calabi-Yau manifolds, mirror symmetry can be understood, in terms
of the underlying conformal field theory description, as the effect of
a non-trivial automorphism of the (right-moving) extended $\cN=2$
superconformal algebra that is always present for Calabi-Yau
compactifications \cite{Odake:1989bh}. The extended algebra for G$_2$
compactifications has been constructed in \cite{Shatashvili:1994zw}
(see also \cite{Figueroa-O'Farrill:1997hm}), and one may ask whether
mirror symmetry for G$_2$ manifolds can be similarly interpreted. The
G$_2$ algebra contains a non-trivial automorphism that leaves the $\cN=1$ 
superconformal subalgebra invariant.\footnote{This automorphism was
already observed in \cite{BBMOOZ}, see also \cite{Roiban:2002iv}.} 
By considering three T-dualities along suitable directions one can
induce this automorphism on the right-movers (without modifying the
left-movers). Depending on which realisation one chooses, this maps
the IIA/IIB theory on one of the nine G$_2$ orbifolds to the IIB/IIA
theory on the {\it  same} orbifold, or to IIB/IIA theory on the
orbifold where {\it all discrete torsion phases have been inverted}. 
In the latter case, the `mirror map' therefore relates the nine G$_2$ 
manifolds pairwise (with one manifold being its own mirror); the
former possibility, on the other hand, is the string theory
realisation of the symmetry proposed in \cite{PT}.
\medskip

This paper is organised as follows. In section~2 we discuss the
generalised orbifold construction in the simpler example of the
Calabi-Yau compactification on $T^6/\Bz_2^2$ for which discrete
torsion and mirror symmetry were studied in detail in
\cite{Vafa:1995rv}. We explain how to solve the constraints imposed by
modular invariance, and show that there are (up to suitable
relabellings) seventeen different theories. Furthermore, we study the
effect of mirror symmetry on these theories, and demonstrate that
these seventeen theories are pairwise identified by mirror
symmetry, with one being its own mirror. From a geometric point of
view, the different theories correspond to the different
desingularisations that can be chosen for the different orbifold
singularities. 

In section~3, the analogous construction is performed for the
$T^7/\Bz_2^3$ orbifold of Joyce (see 
\cite[chapter 12.3]{Joyce:book}). In this case, the analysis of
modular invariance is more cumbersome, and some of the details are
spelled out in the appendix. There are now nine different
orbifold theories, and they correspond precisely to the different
resolutions of Joyce. We also explain how the mirror automorphism of
the G$_2$ manifold can be implemented for this theory, and how it
relates either a G$_2$ manifold to itself, or to a different G$_2$
manifold. Finally, section~4 contains our conclusions.

\section{Generalised discrete torsion and mirror symmetry for 
Ca\-la\-bi-Yau 3-folds} 
\label{cy}

In this section we describe how to generalise discrete torsion for the 
familiar or\-bi\-fold $X=T^6/\Bz_2^2$ that was studied in detail in
\cite{Vafa:1995rv} (see also \cite{Gaberdiel:2000fe}). As we shall
see, the different constructions correspond geometrically to different
choices for how to desingularise the various orbifold singularities.
Finally we shall discuss how certain T-dualities implement the  mirror
symmetry,  and  how this exchanges orbifolds with different choices of
discrete torsion.

\subsection{Generalised discrete torsion}
\label{orbrep}

The Hilbert space $\cH$ of an orbifold theory for an abelian orbifold
group $\Gamma$ consists of sectors $\cH_h$, one such sector for each
element $h$ of the orbifold group $\G$. The sector $\cH_h$ describes
those closed string states that are twisted by the action of $h$ along 
their spacelike direction, \ie\ $x(\tau,\si=2\pi)=h\cdot x(\tau,\si=0)$.  
To be explicit, we consider the $T^6/\Bz_2^2$ orbifold of 
\cite{Vafa:1995rv}, where the two generators of $\G=\Bz_2\times \Bz_2$ 
act multiplicatively as 
\begin{equation}
\begin{array}{r|rrrrrr}
{} & x_1 & x_2 & x_3 & x_4 & x_5 & x_6  \\ \hline
\al & +1 & +1  & -1  & -1  & -1  & -1   \\
\beta & -1 & -1 & +1 & +1  & -1  & -1   \\
\end{array} \label{action.vafa}
\end{equation}
on the coordinates of the torus. We will call $I_h^+$ and $I_h^-$ the
index set of those coordinates on which $h$ acts with even and odd
parity, respectively.

Since each non-trivial orbifold group element inverts four of the six
directions, each has sixteen fixed points. The twisted sector $\cH_h$
can therefore be decomposed into sixteen isomorphic Hilbert spaces
$\cH_{h;f}$, one for each fixed point $f$. 
For each fixed point, 
the Hilbert space $\cH_{h;f}$ is generated by the action of the
oscillators from a ground state that is characterised by its momentum
and winding number. (In the twisted sector $\cH_h$, momentum and
winding numbers only exist for the directions $I^+_h$.) The total
space of states of the orbifold theory consists then of the sum
of all twisted sectors, where in each twisted sector only the
states that are invariant under the action of the orbifold group
survive. The complete partition function of the theory is therefore
given by 
\setlength{\unitlength}{1mm}
\begin{equation}
Z(q,\bq)=\frac{1}{|\G|}\sum_{g,h\in\G} \Tr_{\cH_h}
\left(g\, q^{L_0-c/24}\bq^{\bL_0-c/24}\right) 
=: \frac{1}{|\G|}\sum_{g,h\in\G} 
\begin{picture}(10,10)(0,0)
\put(5,0){\framebox(4,4){}}
\put(7,-3){\mbox{${}_h$}}
\put(2,2){\mbox{${}_g$}}
\end{picture}\,,
\end{equation}
where the sum over $g\in\Gamma$ implements the projection onto
$\Gamma$-invariant states. At first this expression is somewhat formal
since {\it a priori} it is not clear how to define the action of $g$
in the $h$-twisted sector (unless $h=e$, in which case $\cH_e$ is just
the original space of states). 

Suppose now that one has found one consistent action of $g$ in each
$h$-twisted sector that leads to a modular invariant partition
function. The idea of discrete torsion is that one can modify this
action by a phase $\epsilon(g,h)$,
\begin{equation}
\left. \hat{g} \right|_{\cH_h} := \epsilon(g,h) 
\left.  g \right|_{\cH_h}\,,
\end{equation}
where $\epsilon(g,h)$ depends on $g$ and $h$, but is otherwise the
same for {\it all} states in the sector $\cH_h$. In order for the
action defined by $\hat{g}$ to define a representation of $\Gamma$, we
need that  
\begin{equation}\label{repprop}
\epsilon(g_1 \, g_2,h) = \epsilon(g_1,h)\, \epsilon(g_2,h) \,.
\end{equation}
Furthermore, in order for the resulting theory to be modular invariant,
one requires \cite{Vafa}
\begin{equation}
\epsilon(g,h) = \epsilon(g^a h^b,g^c h^d) \,, \qquad ad-bc =1 \,.
\end{equation}
Different sets of discrete torsion phases are in one-to-one
correspondence with elements in $H^2(\Gamma,U(1))$.

In the above, we have modified the action of $g$ on $\cH_h$ by an
overall phase that is the same for all states in the sector
$\cH_h$. Provided that the phases satisfy (\ref{repprop}) this clearly
defines a consistent action of $\Gamma$ on $\cH_h$. However, in
general this is not the only way in which we can modify the action of
$\Gamma$ on $\cH_h$. As we have mentioned above, each $\cH_h$ is the
direct sum of sixteen copies,
\begin{equation}\label{decompf}
\cH_h = \bigoplus_{f=1}^{16} \cH_{h;f} \,,
\end{equation}
where $\cH_{h;f}$ describe the $h$-twisted states associated to the
fixed point $f$. Each of these spaces forms an irreducible
representation of the oscillators. Since $\Gamma$ has a prescribed
action on the oscillators, the action of $g$ on a given state in
$\cH_{h;f}$, determines the action on all of $\cH_{h;f}$ uniquely. On
the other hand, it does {\it not} determine the action of $g$ on the
states in $\cH_{h;f'}$ with $f'\ne f$. Thus, we should be able to
choose discrete torsion phases separately for the different fixed
point components, \ie\ the discrete torsion phase should be allowed to
depend on $f$,
\begin{equation}
\left. \hat{g} \right|_{\cH_{h;f}} := \epsilon_f(g,h) 
\left.  g \right|_{\cH_{h;f}}\,.
\end{equation}
In order for this to define a consistent action of $\Gamma$ on $\cH$,
each $\epsilon_f(g,h)$ must satisfy (\ref{repprop}). In addition, we
must require that the resulting partition function is still modular
invariant. This last condition requires a little bit of care and
depends on the specifics of the theory in question. In order to
analyse this issue, let us write out the contribution of the various
sectors to the partition function (here 
$g,h\in\{\al,\beta,\al\beta\}$ and we have only written down the
bosonic contributions)
\begin{align}
&\begin{picture}(10,10)(0,0)
\put(5,0){\framebox(4,4){}}
\put(7,-3){\mbox{${}_e$}}
\put(2,2){\mbox{${}_e$}}
\end{picture}
=\Tr_{\cH_e} \left( q^{L_0-c/24}\bq^{\bL_0-c/24} \right)
= \frac{1}{|\eta|^{12}}\sum_{(p_L,p_R)\in\G^{6,6}}
q^{\frac{1}{2}p_L^2}\bq^{\frac{1}{2}p_R^2}\, , \non \\
&\begin{picture}(10,10)(0,0)
\put(5,0){\framebox(4,4){}}
\put(7,-3){\mbox{${}_e$}}
\put(2,2){\mbox{${}_g$}}
\end{picture}
=\Tr_{\cH_e} \left(g\, q^{L_0-c/24}\bq^{\bL_0-c/24}\right)
= 16 \frac{1}{|\vth_2|^{4}}\sum_{(p_L,p_R)\in\G^{2,2}}
q^{\frac{1}{2}p_L^2}\bq^{\frac{1}{2}p_R^2}\, , \non \\
&\begin{picture}(10,10)(0,0)
\put(5,0){\framebox(4,4){}}
\put(7,-3){\mbox{${}_g$}}
\put(2,2){\mbox{${}_e$}}
\end{picture}
=\Tr_{\cH_g} \left(q^{L_0-c/24}\bq^{\bL_0-c/24}\right)
= \sum_{f=1}^{16} \frac{1}{|\vth_4|^{4}}\sum_{(p_L,p_R)\in\G^{2,2}}
q^{\frac{1}{2}p_L^2}\bq^{\frac{1}{2}p_R^2}\, , \non \\
&\begin{picture}(10,10)(0,0)
\put(5,0){\framebox(4,4){}}
\put(7,-3){\mbox{${}_g$}}
\put(2,2){\mbox{${}_g$}}
\end{picture}
=\Tr_{\cH_g} \left(g\, q^{L_0-c/24}\bq^{\bL_0-c/24}\right)
= \sum_{f=1}^{16} \eps_f(g,g)
\frac{1}{|\vth_3|^{4}}\sum_{(p_L,p_R)\in\G^{2,2}}
q^{\frac{1}{2}p_L^2}\bq^{\frac{1}{2}p_R^2}\, , \non \\
& \begin{picture}(10,10)(0,0)
\put(5,0){\framebox(4,4){}}
\put(7,-3){\mbox{${}_h$}}
\put(2,2){\mbox{${}_g$}}
\end{picture}|_{g\neq h}
=\left. \Tr_{\cH_h} \left(g\, q^{L_0-c/24}\bq^{\bL_0-c/24}\right)
\right|_{g\neq h}
= \sum_{f=1}^{16} \eps_f(g,h)\, .\non
\end{align} 
Here
\[
\eta=q^{1/24}\prod_{n=1}^{\infty}\left(1-q^n\right)
=\sum_{n=-\infty}^{\infty} (-1)^n 
q^{\frac{3}{2}\left(n-\frac{1}{6}\right)^2}
\]
is the Dedekind $\eta$-function, while the $\vth$-functions are given by
\[
\vth_2=\sum_{n=-\infty}^{\infty} 
q^{\frac{1}{2}\left(n-\frac{1}{2}\right)^2}
\, , \qquad
\vth_3=\sum_{n=-\infty}^{\infty} q^{\frac{1}{2}n^2}
\, , \qquad
\vth_4=\sum_{n=-\infty}^{\infty} (-1)^n q^{\frac{1}{2}n^2}\,.
\]
The lattice $\Gamma^{6,6}$ is the full momentum and winding lattice
of the underlying $T^6$, which we assume to decompose as 
$T^6=T^2\times T^2 \times T^2$. As a consequence, $\Gamma^{6,6}$ is a 
direct sum of three $\Gamma^{2,2}$ lattices that are associated to the
three different $T^2$s. All of these lattices are even and self-dual,
and the corresponding partition functions therefore transform
in a simple manner under the modular group. The modular invariance of
the full partition function then requires that 
\begin{align}
&\begin{picture}(10,10)(0,0)
\put(5,0){\framebox(4,4){}}
\put(7,-3){\mbox{${}_g$}}
\put(2,2){\mbox{${}_e$}}
\end{picture}(q(\tau+1))=
\begin{picture}(10,10)(0,0)
\put(5,0){\framebox(4,4){}}
\put(7,-3){\mbox{${}_g$}}
\put(2,2){\mbox{${}_g$}}
\end{picture}(q(\tau))
\quad &&\Rightarrow \quad
\eps_f(g,g)=1 , \quad \forall ~g\,,f\,, \label{mod1} \\
&\begin{picture}(10,10)(0,0)
\put(5,0){\framebox(4,4){}}
\put(7,-3){\mbox{${}_h$}}
\put(2,2){\mbox{${}_g$}}
\end{picture}(q(-1/\tau))=
\begin{picture}(10,10)(0,0)
\put(5,0){\framebox(4,4){}}
\put(7,-3){\mbox{${}_g$}}
\put(2,2){\mbox{${}_h$}}
\end{picture}(q(\tau))
\quad &&\Rightarrow \quad
\sum_{f=1}^{16}\eps_f(g,h)=\sum_{f=1}^{16}\eps_f(h,g)\, . \label{mod2}
\end{align}
The first of these constraints (that arises from the modular 
$T$-transformation) leaves us with 16 signs 
$\eps_{h;f}$, $f=1,\ldots,16$ for each twisted sector labelled by $h$.
[For example,
$\eps_{\alpha;f} = \eps_f(\beta,\alpha)=\eps_f(\alpha\beta,\alpha)$.]
The second constraint (that arises from the modular $S$-transformation)
forces the number of positive signs to be the same in each sector. 
By relabelling the fixed points if necessary, we can therefore set 
$\eps_f:= \eps_{\al;f}=\eps_{\beta;f}=\eps_{\al\beta;f}$.

In \cite{Vafa} it was shown that modular invariance at one-loop, \ie\  
(\ref{mod1}) and (\ref{mod2}), together with (\ref{repprop})
imply via the factorisation property that the (bosonic) orbifold is 
modular invariant on any genus $n$ Riemann surface (at least in the
limit where the latter degenerates into $n$ 1-tori connected by
infinitly long and thin cylinders). For generalised discrete torsion
the analogeous argument does not work any more; this is to say, the
one-loop modular invariance (together with the factorisation property)
does not imply automatically that the theory is modular invariant on
higher genus surfaces as well. [The problem arises when trying to
show that the vacuum amplitude is invariant under the Dehn twist that
links adjacent tori.] On the other hand, this does not imply that the
theory with generalised discrete torsion does not satisfy higher genus
modular invariance; it merely means that higher genus modular 
invariance will only hold provided that the vacuum amplitudes 
$A(g_1,h_1,f_1;g_2,h_2,f_2;\ldots;g_n,h_n,f_n)$ satisfy suitable
additional properties. [For example, one can show that 2-loop modular
invariance follows for suitable assignments of the different discrete
torsion phases to the different fixed points provided that the
amplitudes $A(g_1,h_1,f_1;g_2,h_2,f_2;\ldots;g_n,h_n,f_n)$ involving
`different' fixed points vanish in the above limit.] It would
obviously be very interesting to analyse this question for the 
theories we discuss here, but unfortunately, this is a very difficult
problem which seems to be out of reach at the moment. [It would
require constructing the actual 2-loop amplitudes for the various
twisted sectors and twists, but such amplitudes have not even be
constructed in much simpler examples.] It is therefore conceivable
that additional restrictions on the choice of the different discrete
torsion phases will be required by higher genus modular invariance. 

The contribution of the
fermions is also described by $\vth$ functions, and their inclusions
does not destroy the modular invariance properties.\footnote{Strictly
speaking, the inclusion of fermions introduces fermionic zero modes
into various sectors which in turn make the associated vacuum amplitudes
vanish. Formally, the constraints of modular invariance therefore seem
to be weaker in the fermionic case. 
However, these `accidental'
vanishings can be lifted by considering torus amplitudes that include
an appropriate number of fermionic zero modes, and it is thus believed
that the bosonic conditions are necessary and sufficient to guarantee
modular invariance in the fermionic case as well. The inclusion of
fermions may on the other hand modify the conditions for modular
invariance at higher genus.}

In total there are therefore 17 different choices of discrete torsion 
given by the number $\ell\in\{0,\ldots,16\}$ of positive signs among
$\eps_f$.\footnote{Actually, the number of different theories is 
bigger since theories that differ in the way these signs are 
qdistributed among the fixed points will in general be different. However,
their Betti/Hodge numbers will be the same.}
The extremal cases $\ell=0$ and $\ell=16$ correspond
precisely to the situation with and without discrete torsion,
respectively \cite{Vafa:1995rv}, but now we also have intermediate
possibilities.

\subsection{Discrete torsion and the resolution of orbifold
singularities} 

In this section we will give a geometrical interpretation of these
generalised discrete torsion theories. 
It is common lore that the untwisted sector captures the
geometry of the singular orbifold, whereas the twisted sectors
describe their resolution. It is therefore not surprising that
discrete torsion has to do with the way in which one resolves the
singularities. We will find that the spectrum of ground states in the 
$h$-twisted sector depends on discrete torsion if the singularity of
$h$-fixed loci can be resolved in inequivalent ways. A particular
choice of discrete torsion then tells us which resolution is
chosen. Our generalisation of discrete torsion corresponds thus simply
to the possibility of choosing different resolutions for different
fixed points.

In order to relate the orbifold CFT to the topology of the 
target space we exploit the isomorphism between the space of 
RR ground states and the cohomology of the target space 
\cite{Witten:1982im}. To this end
we accompany the left- and right-moving part of each coordinate $x_j$
with a left- and right-moving (2d) Majorana-Weyl spinor $\psi^j$ and
$\tpsi^j$, respectively. If the original metric on $T^6$ was chosen to 
be the flat one, their zero modes satisfy the Clifford algebra
\begin{equation}
\{\psi_0^i,\psi_0^j\}=2\delta^{ij} \,, \quad 
\{\tpsi_0^i,\tpsi_0^j\}=2\delta^{ij} \,, \quad 
\{\psi_0^i,\tpsi_0^j\}=0 \,. \label{cliff1}
\end{equation} 
In order to build the Fock space of physical states we define
\begin{equation}
\psi^j_{\pm}:=\frac{1}{2}\left(\psi^j_0\pm i\tpsi_0^j\right) \,,
\qquad j=1,\dots,6 \label{psipm}
\end{equation}
which satisfy the algebra
\begin{equation}
\{\psi^i_{\pm},\psi^j_{\mp}\}=\delta^{ij} \,, \quad 
\{\psi^i_{\pm},\psi^j_{\pm}\}=0 \,. \label{cliff2}
\end{equation} 
We can then choose the $\psi^j_+$ to be creators and $\psi^j_-$ to be 
annihilators,
\begin{equation}
\psi^j_- |0\rangle =0 
\qquad \Rightarrow \qquad
\psi_0^j |0\rangle = i \tpsi_0^j |0\rangle \label{create}
\end{equation}
in the Fock space. Note that the above is not the standard choice of 
generators for \CY target spaces (see e.g.\ \cite{Greene:1996cy});
however, the above convention makes also sense for spaces of odd
dimension (such as the G$_2$ spaces we shall consider in the next
section) and it is therefore more convenient for us than the usual
definition which only works for spaces of even dimension. Of course
the two choices generate isomorphic Fock spaces.

The orbifold invariant RR ground states in the untwisted sector are
then given by 
\begin{align}
&|0\rangle \non \\
&\psi^1_+ \psi^2_+ |0\rangle , ~~
\psi^3_+ \psi^4_+ |0\rangle , ~~
\psi^5_+ \psi^6_+ |0\rangle  \non \\
& \psi^1_+ \psi^3_+ \psi^5_+|0\rangle , ~~
\psi^1_+ \psi^3_+ \psi^6_+|0\rangle , ~~
\psi^1_+ \psi^4_+ \psi^5_+|0\rangle , ~~
\psi^1_+ \psi^4_+ \psi^6_+|0\rangle \non \\
& \qquad \qquad
\psi^2_+ \psi^3_+ \psi^5_+|0\rangle , ~~
\psi^2_+ \psi^3_+ \psi^6_+|0\rangle , ~~
\psi^2_+ \psi^4_+ \psi^5_+|0\rangle , ~~
\psi^2_+ \psi^4_+ \psi^6_+|0\rangle \non \\
&\psi^1_+ \psi^2_+ \psi^3_+ \psi^4_+ |0\rangle , ~~
\psi^1_+ \psi^2_+ \psi^5_+ \psi^6_+ |0\rangle , ~~
\psi^3_+ \psi^4_+ \psi^5_+ \psi^6_+ |0\rangle  \non \\
&\psi^1_+ \psi^2_+ \psi^3_+ \psi^4_+ \psi^5_+ \psi^6_+ |0\rangle \,.
\end{align} 
Identifying
\[
\psi^{j_1}_+ \ldots \psi^{j_n}_+ |0\rangle 
\simeq dx^{j_1} \wedge \ldots \wedge dx^{j_n} \label{ident}
\] 
these RR ground states become the $\G$-invariant harmonic forms on $T^6$.

Since the three twisted sectors are isomorphic we will discuss the 
RR ground states of only one of them, say the $\al$-twisted sector.
In the RR sector then only $\psi^1$ and $\psi^2$ have zero-modes
since the twist changes the spin structure of the remaining fermions.  
Let $|0,0;f\rangle_{\al}$ be one of the 16 highest weight states with
vanishing momentum and winding. The $\G$-invariant RR ground states built 
on this state are then 
\begin{equation}
|0,0;f\rangle_{\al} \qquad \mbox{and} \qquad
\psi^1_+ \psi^2_+ |0,0;f\rangle_{\al}
\end{equation}
if $\beta|0,0;f\rangle_{\al}=\alpha\beta |0,0;f\rangle_{\al} 
= |0,0;f\rangle_{\al}$, \ie\
$\eps_{\al;f}=1$, or  
\begin{equation}
\psi^1_+ |0,0;f\rangle_{\al} \qquad \mbox{and} \qquad
\psi^2_+ |0,0;f\rangle_{\al}
\end{equation}
if $\beta|0,0;f\rangle_{\al}=\alpha\beta |0,0;f\rangle_{\al} 
=-|0,0;f\rangle_{\al}$, \ie\
$\eps_{\al;f}=-1$. Again we can identify these states with harmonic 
forms. The state $|0,0;f\rangle_{\al}$ however is now not identified
with the constant zero-form but with the two-form $\om_{\al;f}$
representing the class of the exceptional divisor that arises in the
resolution of this singularity. In the complex structure  
\[
z_1=x_1+i x_2\, ,\qquad 
z_2=x_3+i x_4\, ,\qquad 
z_3=x_5+i x_6\, ,
\]
this form will be of type (1,1). Using the same complex structure to
complexify the fermions, the highest weight sector
$|0,0;f\rangle_{\al}$ contributes one class to $h^{1,1}$ and $h^{2,2}$
for $\eps_{\al;f}=1$, and one class to $h^{2,1}$ and $h^{1,2}$ for
$\eps_{\al;f}=-1$. 

Let $\ell\in\{0,\ldots,16\}$ be the number of positive signs among
$\eps_f$, then since all three twisted sectors are isomorphic their
total contribution to the cohomology of the target space is 
\begin{equation}
h^{p,q}(\mbox{twisted sectors})=\begin{pmatrix}
0 & 0 & 0 & 0 \\
0 & 3\ell & 48-3\ell & 0 \\
0 & 48-3\ell & 3\ell & 0 \\
0 & 0 & 0 & 0 \end{pmatrix} ,
\end{equation}
where $p$ and $q$ label the rows and columns respectively.
In total the 17 different choices of discrete torsion lead to target 
manifolds $X_{\ell}$ with Hodge diamonds
\begin{equation}
h^{p,q}(X_{\ell})=\begin{pmatrix}
1 & 0 & 0 & 1 \\
0 & 3(\ell+1) & 51-3\ell & 0 \\
0 & 51-3\ell & 3(\ell+1) & 0 \\
1 & 0 & 0 & 1 \end{pmatrix} , \quad \mbox{for}~\ell\in\{0,\ldots,16\} .
\label{cohx}
\end{equation}

Next we will show that the choice of the discrete torsion phase at a
fixed point is associated with the choice of how the corresponding
singularity is resolved.
Consider for definiteness the complex codimension two 
singularities due to the action of $\al$.   
Locally each of these 16 singularities inside $T^6/\G$ looks like
\begin{equation}
(T^2\times \Bc^2/\{\pm 1\})/\langle \beta,\al\beta\rangle . 
\end{equation}
As is explained in \cite{Joyce:book}, there are two inequivalent ways to
desingularise the $\Bc^2/\{\pm 1\}$ singularity inside the total
orbifold.

The first one is the blow up $Y_1$ of 
$\Bc^2/\{\pm 1\}=(z_2,z_3)/\{\pm 1\}$ at the origin,
which creates an exceptional divisor 
$\Si_1=\Bc P^1\simeq [z_2,z_3] \subset Y_1$ whose
homology class generates $H_2(Y_1,\Br)=\Br$.   
The actions of $\beta$ and $\al\beta$ lift to $\Si_1$ and act as
\begin{align}
\beta &: [z_2,z_3] \mapsto [z_2,-z_3] \,, \non \\
\al\beta &: [z_2,z_3] \mapsto [-z_2,z_3] \,, \non 
\end{align}
which both preserve the orientation of $\Si_1$ and hence the induced 
maps $\beta_*$ and $\al\beta_*$ on $H_2(Y_1,\Br)$ are the identity.

The second way to desingularise $\Bc^2/\{\pm 1\}$
is to deform it. To this end one defines 
$\si~:~\Bc^2/\{\pm 1\}\rightarrow \Bc^3$ by
\begin{equation}
\si~:~ \pm (z_2,z_3)\mapsto 
(z_2^2-z_3^2,iz_2^2+iz_3^2,2z_2z_3)\, ,
\end{equation}
which identifies $\Bc^2/\{\pm 1\}$ with the quadratic
\[
\{(w_1,w_2,w_3)\in\Bc^3~|~w_1^2+w_2^2+w_3^2=0 \} \,.
\]
Let $\eta\in\Bc^{\times}$ be small and non-zero and
\begin{equation}
Y_2:=\{(w_1,w_2,w_3)\in\Bc^3~|~w_1^2+w_2^2+w_3^2=\eta \} \,,
\end{equation}
then $Y_2$ is a smoothing of $\Bc^2/\{\pm 1\}$ that is diffeomorphic
to $Y_1$. Let $\eta=re^{2i\phi}$ with $r\in \Br$ positive and 
$\phi\in [0,\pi)$, and define
\begin{equation}
\Si_2:=\{(e^{i\phi}x_1,e^{i\phi}x_2,e^{i\phi}x_3)~|~x_j\in\Br,~
x_1^2+x_2^2+x_3^2=r\} \,,
\end{equation}
then the homology class of $\Si_2\simeq S^2\subset Y_2$ generates
$H_2(Y_2,\Br)=\Br$. Since $\beta$ and $\al\beta$ both act as
\[
\beta,\al\beta~:~(w_1,w_2,w_3)\mapsto (w_1,w_2,-w_3) \,,
\] 
they preserve  $\Si_2$ but reverse its orientation. Thus their induced
maps $\beta_*$ and $\al\beta_*$ on $H_2(Y_2,\Br)$ are minus the identity.

Since the exceptional divisor $\Si_{\al;f}$ (generating 
$H_2(Y_{1},\Br)$ or $H_2(Y_{2},\Br)$, respectively) corresponds  
to the ground state $|0,0;f\rangle_{\al}$
of the corresponding twisted sector, the discrete torsion signs 
$\eps_{\al;f}$ are geometrically just the eigenvalues 
of $\beta_*$ (and $\al\beta_*$) on the homology classes of the
exceptional divisors $\Si_{\al;f}$. 
[The fact that these eigenvalues appear in the partition function is
due to the contribution of the $B$-field evaluated on these 
classes; for a geometrical description of discrete torsion as a choice
of representation of the orbifold group on the $B$-field see 
\cite{Sharpe:2000ki}.] 
Hence
the parameter $\ell$ counts how many of the 16 singularities 
of complex codimension two generated by $\al$ we have chosen
to blow up, instead of deforming it. The analysis is obviously
identical for the sectors twisted by $\beta$ and $\alpha\beta$. The
cohomology of the resulting space is then exactly the one given in
(\ref{cohx}). 

{}From this geometric point of view, it is {\it a priori} not clear
why $\ell$ has to be taken to be the same for all three twisted
sectors. (This is the condition that arose from the requirement that
the partition function is invariant under the modular
$S$-transformation.) However, the reason may be related to the fact
that the orbifold has an $S_3$-symmetry of permutations of  
$z_1,z_2,z_3$ that also permutes the three twisted sectors. If this
symmetry is to be respected by the discrete torsion phases, then
$\ell$ indeed has to be the same for all three twisted sectors.

\subsection{Discrete torsion and mirror symmetry of orbifold 
Calabi-Yau 3-folds}

Finally we want to show that, in this example, 
T-duality on three of the coordinates generates 
mirror symmetry and that this exchanges the orbifold with discrete torsion 
parameter $\ell$ with the one with parameter $16-\ell$ in
which all the discrete torsion signs are reversed. This is a
generalisation of the result of \cite{Vafa:1995rv}, where this was
shown by other means for $\ell=0$.

The chiral algebra of a string on a \CY 3-fold
\cite{Odake:1989bh} consists of the
$\cN=2$ superconformal algebra generated by the stress energy tensor
$T_{\rm CY}$, the two supercurrents $G_{\rm CY}, G^{\prime}_{\rm CY}$ and
the U(1)-current $J$ that is extended
by a complex current $\Om$ of conformal weight
$h_{\Om}=3/2$ and its superpartner $\Psi:=\{G_{\rm CY},\Om\}$. 
For the above orbifold theory and with our choice of  
complex structure, the relevant currents look like
\begin{align}
T_{\rm CY}&= \frac{1}{2}\sum_{j=1}^6 :\partial x_j \partial x_j : 
-\frac{1}{2} \sum_{j=1}^6 : \psi^j \partial \psi^j :  \, , \qquad
G_{\rm CY}= \sum_{j=1}^6 : \psi^j \partial x_j : \,, \non \\
G^{\prime}_{\rm CY}&=
\sum_{j=1}^3
\left(\psi^{2j-1}\partial x_{2j}-\psi^{2j}\partial x_{2j-1}\right) 
\,, \qquad
J= 
\sum_{j=1}^3 \psi^{2j-1}\psi^{2j}  \non \\
\intertext{and}
\Om&=\psi^1\psi^3\psi^5 -\psi^1\psi^4\psi^6
-\psi^2\psi^3\psi^6 -\psi^2\psi^4\psi^5 \non \\
&\qquad\qquad+i\left(\psi^1\psi^3\psi^6 +\psi^1\psi^4\psi^5
+\psi^2\psi^3\psi^5 -\psi^2\psi^4\psi^6\right) \,, \label{Om}
\end{align}
which are all preserved by the orbifold action (\ref{action.vafa}).
There is an isomorphic right-moving chiral algebra in which all 
fields/operators are replaced by their right-moving partners, e.g.\ 
$\psi^j\rightarrow \tpsi^j$.

This chiral algebra has two interesting automorphisms 
\cite{Figueroa-O'Farrill:1997hm}. 
The first one is
a simultaneous phase rotation of the extension operators
\begin{equation}
\Om\mapsto e^{i\phi}\Om 
\qquad \mbox{and}\qquad
\Psi \mapsto e^{i\phi}\Psi \,. \label{aut1}
\end{equation}
The second one is the mirror automorphism
\begin{equation}
\mbox{mirror}_{\rm CY}~:~
G^{\prime}_{\rm CY}\mapsto -G^{\prime}_{\rm CY}\, ,\quad
J\mapsto -J \,, \quad
\Om \mapsto \Om^* \,, \quad
\Psi \mapsto \Psi^* \,, \label{aut2}
\end{equation}
with the remaining operators (in particular the $\cN=1$ superconformal 
subalgebra spanned by $T_{\rm CY}$ and $G_{\rm CY}$) being invariant.

Mirror symmetry for \CY manifolds was discovered \cite{Lerche:1989uy} as a 
consequence of applying the second automorphism to one of the chiral
algebras, say the right-moving one (with tildes). 
Using the free field representation of the algebra operators, and combining
the two automorphisms (with the phase of the former being equal to
$e^{i\phi}=\pm 1$) we see that T-duality on three coordinates
$x_{j_1},x_{j_2},x_{j_3}$ with 
\begin{equation}
(j_1,j_2,j_3)\in \{ (1,3,5),(1,4,6),(2,3,6),(2,4,5),
(1,3,6),(1,4,5),(2,3,5),(2,4,6) \} \label{j3}
\end{equation} 
generates the mirror automorphism
on the right-moving chiral algebra while leaving the left-moving one
invariant. [Note that the combinations of labels are precisely those that 
appear in the eight terms of $\Om$, (\ref{Om}).] This works because  
T-duality on $x_j$ leaves the left-moving
current $\partial x_j$ invariant, but reverses the right-moving one
$\overline{\partial}x_j$. The same then holds for the worldsheet 
superpartners, \ie\ $\psi^j\mapsto \psi^j$ but 
$\tpsi^j\mapsto -\tpsi^j$. Hence T-duality on these three coordinates
relates mirror manifolds to one other.

Next we want to show that T-duality on these three coordinates 
relates target spaces to each other that correspond to orbifolds where
all the discrete torsion signs are reversed.
To this end consider the $h$-twisted sector
in the RR sector where only the fermions $\psi^j$ and $\tpsi^j$ with
$j\in I_h^+$ have zero-modes. For definiteness, let us take 
$h=\alpha$, for which $I_\al^+=\{1,2\}$; the analysis for the other 
two cases is identical. We are interested in how the action of $\beta$
is modified by the T-duality transformation. Since $\beta$ acts as 
\begin{equation}
\beta \psi^i_0 \beta = - \psi^i_0 \,, \qquad
\beta \tpsi^i_0 \beta = - \tpsi^i_0 \,, \qquad i\in I_\alpha^+=\{1,2\} 
\end{equation}
on the fermionic zero modes, 
we can represent it on the RR ground states of $\cH_{\al;f}$ as 
\begin{equation}
\beta = \frac{1}{4} \psi^1_0 \psi^2_0 \tpsi^1_0 \tpsi^2_0 \cdot
\eps_{\al;f} \,.
\end{equation}
Under any of the above T-duality transformations in (\ref{j3}), 
$\beta$ then changes sign. 
Since the above analysis applies uniformly for 
all fixed points, this operation 
therefore corresponds to changing 
all the $\eps_{\al;f}$. 
Thus we conclude that mirror symmetry inverts all the discrete torsion
signs in this case; in particular, it therefore relates the orbifold 
labelled by $\ell$, $X_\ell$,  to that labelled by $16-\ell$, 
$X_{16-\ell}$. On the Hodge numbers (\ref{cohx}) this generates 
indeed the correct symmetry since 
$h_{1,1}(X_{\ell})=h_{2,1}(X_{16-\ell})$.

\section{Generalised discrete torsion and mirror symmetry for  
G$_{\mathbf 2}$ orbifolds}
\label{g2}

In this section we will apply the same reasoning to the orbifold 
$T^7/\Bz_2^3$ of \cite{Joyce:book,Shatashvili:1994zw} associated to 
Riemannian manifolds of 
holonomy G$_2$. Conceptually everything is as in the previous section.
However, an interesting application will be mirror symmetry for G$_2$ 
manifolds that we consider in section \ref{g2.mirror}.
But before doing so, let us present the model that we are going to work 
with.

\subsection{Compact orbifold G$_{\mathbf 2}$ manifolds and discrete torsion}

Consider the orbifold of (\cite{Joyce:book}, chapter 12.3),
\begin{equation}
Y=T^7/\Bz_2^3 \, , \label{g2orb1}
\end{equation}
where $\G\equiv \Bz_2\times \Bz_2\times \Bz_2$ is generated by
\begin{equation}
\begin{array}{l@{~\equiv ~[(}r@{,}r@{,}r@{,}r@{,}r@{,}r@{,}r@{)~;~(}
c@{,~}c@{,~}c@{,~}c@{,~}c@{,~}c@{,~}c}
\al & -1 & -1 & -1 & -1 & 1 & 1 & 1 &
 0  & 0  & 0  & 0  & 0  & 0 & 0 \, )]\, , \\ 
\beta & -1 & -1 & 1 & 1 & -1 & -1 & 1 &
 0  & 1/2 & 0  & 0  & 0  & 0 & 0 \, )]\, , \\ 
\ga & -1 & 1 & -1 & 1 & -1 & 1 & -1 &
 1/2 & 0  & 0  & 0  & 0  & 0 & 0 \, )]\, . \\ 
\end{array}  \label{g2orb2}
\end{equation}
Here the entries in the first vector denote the eigenvalue of the
coordinates $x_j$ under the multiplicative group action, 
$x_j\mapsto \pm x_j$, and the entries of the 
second vector denote shifts $x_j\mapsto x_j+\eps$. 
The action of $\ga$ on $x_1$ for example is $\ga : x_1\mapsto -x_1+1/2$.
Moreover, the coordinates are taken to have unit periodicity,
$x_j\equiv x_j+ 1$.  

The elements $\al\beta,\al\ga,\beta\ga,\al\beta\ga$ of $\G$ have no
fixed  points on $T^7$ due to the shifts in the first or second
coordinate.  
The fixed points of $\al,\beta,\ga$ in $T^7$ are each 16 copies of $T^3$,
where $\langle \beta,\ga\rangle$ acts freely on the 16 $\al$-fixed $T^3$s 
and $\langle \al,\ga\rangle$ acts freely on the 16 $\beta$-fixed $T^3$s,
building four orbits of four tori in each case. However, 
$\langle \al,\beta\rangle$ does not act freely on the set of 16 
$\ga$-fixed tori, since $\al\beta$ acts trivially (while $\al$ and $\beta$
build eight orbits of order two).
The singular set of $T^7/\G$ is thus a disjoint union of eight copies
of $T^3$ and eight copies of $T^3/\Bz_2$. The singularity at each $T^3$ 
is locally modelled on $T^3\times \Bc^2/\{\pm 1\}$ whereas the one at
each $T^3/\Bz_2$ is modelled on 
$(T^3\times \Bc^2/\{\pm 1\})/\langle \al\beta\rangle$. 
The resolution of the latter is not unique in the same way as in the \CY 
case, due to the different action of $\al\beta_*$ on the exceptional
divisors arising from the blow up or the deformation of $\Bc^2/\{\pm 1\}$.
In fact the analysis can be taken over word by word from the previous
section. If $\ell\in\{0,\ldots,8\}$ denotes the number of $T^3/\Bz_2$ 
singularities that we choose to blow up, then we generate 
nine topologically different manifolds $Y_{\ell}$ with Betti-numbers
\begin{equation}
(b_0,\ldots,b_7)(Y_{\ell})=(1,0,8+\ell,47-\ell,47-\ell,8+\ell,0,1)
,\quad \mbox{for}~\ell=0,1,\ldots,8 . \label{bjoyce}
\end{equation}
Joyce \cite{Joyce:book} has shown 
that all of these are compact G$_2$ manifolds.

Next we will discuss how the above nine classes of 
G$_2$ manifolds are in one-to-one correspondence with nine choices of 
generalised discrete torsion for this orbifold. As before, we
introduce discrete torsion phases for the various twisted sectors, and
allow them to be different for the different fixed 
points or twist fields.\footnote{Strictly speaking, 
for each nontrivial $h\in\G$, the different subspaces for which
separate discrete torsion phases can be introduced are not labelled by 
the fixed points of the action of $h$, but by the twist fields of
lowest conformal dimension that generate the irreducible
representations of the oscillators as in (\ref{decompf}). In our
orbifold there are 16 such fields for any nontrivial $h$, even if $h$
does not have any fixed points.} 
For each $f$ and $h$, the phases $\eps_f(g,h)$ must form a
representation of $\Gamma$ w.r.t.\ the first argument. 
Furthermore, some of these phases are spurious in
that they can be absorbed into the normalisation of the different
states. This analysis is discussed in detail in appendix \ref{app1}. 
After these considerations have been taken into account, we are left
with eight signs  $\eps_{\gamma;\tf}=\pm 1$ for $\tf=1,\ldots,8$, in
the $\ga$-twisted sector; another eight signs 
$\eps_{\al\beta\ga;\tf}=\pm 1$ for $\tf=1,\ldots,8$, in the
$\al\beta\ga$-twisted sector; and sixteen signs
$\eps_{\al\beta;\tf}=\pm 1$ for $\tf=1,\ldots,16$, 
in the $\al\beta$-twisted sector. However, as in the \CY case  
the constraint from the modular $S$-transformation relates their sums
to each other
\begin{equation}
2\sum_{\tf=1}^8 \eps_{\ga;\tf}
=\sum_{\tf=1}^{16}\eps_{\al\beta;\tf}
=2\sum_{\tf=1}^8 \eps_{\al\beta\ga;\tf}\, . \label{Sg2}
\end{equation}
By relabelling the twist fields if necessary we can therefore set 
$\eps_{\ga;\tf}=\eps_{\al\beta\ga;\tf}=\eps_{\al\beta;\tf}
=\eps_{\al\beta;\tf+8}$ for $\tf=1,\ldots,8$. Up to the ambiguity in
how to distribute the various signs among the different twist fields,
there are therefore nine different theories which are parametrised by
the number $\ell\in\{0,\ldots,8\}$ of plus signs among the eight signs 
$\eps_{\ga;\tf}$. 
\smallskip

Next we want to show that these nine choices of discrete torsion
correspond indeed to the nine topological classes of G$_2$ manifolds
(\ref{bjoyce}). For this we have to look at the RR ground states. Such
lowest energy states exist only in the sectors
$\cH_e,\cH_{\al},\cH_{\beta}$ and $\cH_{\ga}$ since in the remaining
sectors one winding mode takes values in $\Bz+\frac{1}{2}$. 

Using the same definitions as in (\ref{cliff1})--(\ref{create}), 
the $\Gamma$-invariant RR ground states in the untwisted sector
$\cH_e$ are 
\begin{equation}
\vert 0\rangle,~ \psi_+^{j_1}\psi_+^{j_2}\psi_+^{j_3} \vert 0\rangle,~ 
\psi_+^{j_1}\psi_+^{j_2}\psi_+^{j_3}\psi_+^{j_4} \vert 0\rangle,~ 
\mbox{and}~ 
\psi_+^{1}\ldots\psi_+^{7}\vert 0\rangle ,  
\end{equation}
where the 3-tupel and 4-tupel of indices take values in 
\begin{align}
(j_1,j_2,j_3)\in &\{(1,3,6),(1,4,5),(2,3,5),(2,4,6),(1,2,7),(3,4,7),
(5,6,7)\} \,, \non \\
(j_1,j_2,j_3,j_4)\in &\{(2,4,5,7),(2,3,6,7),(1,4,6,7),(1,3,5,7),
(3,4,5,6),(1,2,5,6), \non \\
& ~~(1,2,3,4)\} \,. \non
\end{align}
Hence upon the identification (\ref{ident}), the untwisted sector
contributes just the $\G$-invariant harmonic forms of $T^7$, giving
rise to the Betti numbers
\begin{equation}
(b_0^{(e)},\ldots,b_7^{(e)})=(1,0,0,7,7,0,0,1)\,. \label{bnull}
\end{equation}
Next we need to analyse the contributions from the twisted
sectors. Since only the sectors $h=\al,\beta,\ga$ have fixed points,
only the $h$-twisted sectors with $h=\al,\beta,\ga$ give rise to
massless states. Let us first consider the 
$h$-twisted sectors with $h\in\{\al,\beta\}$. The 
action of $\G$ groups the 16 highest weight states $|0,0;f\rangle_h$
with $f=1,\ldots,16$  into 4 independent $\G$-invariant linear
combinations $|0,0;\hat{f}\rangle_h$ for
$\hat{f}=1,\ldots,4$. Geometrically they correspond to 
the exceptional divisors that resolve the four 
$T^3\times\Bc^2/\{\pm 1\}$ singularities that the action of $h$
produces  in $T^7/\G$. The RR ground states in $\cH_h$ are then 
\begin{align}
& |0,0;\hat{f}\rangle_h , \non \\ 
& \psi_+^{j_1} |0,0;\hat{f}\rangle_h \,, \quad
\psi_+^{j_2} |0,0;\hat{f}\rangle_h \,, \quad
\psi_+^{j_3} |0,0;\hat{f}\rangle_h \,, \non \\
& \psi_+^{j_1} \psi_+^{j_2} |0,0;\hat{f}\rangle_h \,, \quad
\psi_+^{j_1} \psi_+^{j_3} |0,0;\hat{f}\rangle_h\,, \quad
\psi_+^{j_2} \psi_+^{j_3} |0,0;\hat{f}\rangle_h\,, \non \\
& \psi_+^{j_1} \psi_+^{j_2} \psi_+^{j_3} |0,0;\hat{f}\rangle_h\,, \non 
\end{align}
where $\hat{f}=1,\ldots,4$ and $(j_1,j_2,j_3)=(5,6,7)$ or 
$(3,4,7)$ for $h=\al$ or $\beta$, respectively, label the untwisted 
directions $j_i\in I_h^+$. 
Provided we choose appropriate relative signs between the four fixed
points that correspond to each $\hat{f}$, all of these states are
invariant under $\Gamma$.
Identifying $|0,0;\hat{f}\rangle_h$ 
with the harmonic two-form $\om_{h;\hat{f}}$ of the exceptional divisor 
$\Si_{h;\hat{f}}$, these ground states correspond to the harmonic
forms on the $h$-fixed $T^3$ wedge $\om_{h;\hat{f}}$, and they
contribute the Betti numbers  
\begin{equation}
(b_0^{(h)},\ldots,b_7^{(h)})=(0,0,4,12,12,4,0,0) , 
\qquad \mbox{for}~h=\al,\beta . \label{bh}
\end{equation} 
This leaves us with the (more interesting) $\ga$-twisted sector. Here
the action of $\G$ groups the 16 highest weight states 
$|0,0;f\rangle_{\ga}$ with $f=1,\ldots,16$ 
into 8 independent $\G$-invariant linear combinations
$|0,0;\tf\rangle_{\ga}$ for $\tf=1,\ldots,8$. These correspond
geometrically to the exceptional divisors that resolve the eight  
$(T^3\times\Bc^2/\{\pm 1\})/\langle \al\beta\rangle$ singularities
that the  action of $\ga$ produces in $T^7/\G$. The $\al\beta$-parity
of $|0,0;\tf\rangle_{\ga}$ is given by the discrete torsion sign  
$\eps_{\ga;\tf}$, and can be chosen to be $\pm 1$ independently for
each of the eight $\tf$. Geometrically this corresponds to blowing up
the $\Bc^2/\{\pm 1\}$ singularity or deforming it, respectively. 
If we choose $\eps_{\ga;\tf}=1$ (the blow up), then this  
singularity contributes the RR ground states
\begin{equation} \label{states}
|0,0;\tf\rangle_{\ga} \,, \quad
\psi_+^{2} |0,0;\tf\rangle_{\ga} \,, \quad
\psi_+^{4} \psi_+^{6} |0,0;\tf\rangle_{\ga} \quad \mbox{and}\quad
\psi_+^{2} \psi_+^{4} \psi_+^{6} |0,0;\tf\rangle_{\ga}   
\end{equation}
and Betti numbers
\begin{equation}
(b_0^{(\ga)},\ldots,b_7^{(\ga)})_{\tf}=(0,0,1,1,1,1,0,0)
\,. \label{bga1} 
\end{equation} 
[Again, provided we choose the appropriate relative sign between the
two fixed points corresponding to a given $\tilde{f}$, all of the
states in (\ref{states}) are invariant under the whole orbifold group
$\Gamma$.] 
However, if we choose $\eps_{\ga;\tf}=-1$ (the deformation), then this  
singularity contributes the RR ground states
\begin{equation}
\psi_+^{4} |0,0;\tf\rangle_{\ga} \,, \quad
\psi_+^{6} |0,0;\tf\rangle_{\ga} \,, \quad
\psi_+^{2} \psi_+^{4} |0,0;\tf\rangle_{\ga} \quad \mbox{and}\quad
\psi_+^{2} \psi_+^{6} |0,0;\tf\rangle_{\ga}   
\end{equation}
and Betti numbers
\begin{equation}
(b_0^{(\ga)},\ldots,b_7^{(\ga)})_{\tf}=(0,0,0,2,2,0,0,0)
\,. \label{bga2} 
\end{equation} 
If we denote by $\ell\in\{0,\ldots,8\}$ the number of positive signs 
among the $\eps_{\ga;\tf}$ then summing up 
(\ref{bnull}), (\ref{bh}), (\ref{bga1}) and (\ref{bga2}) gives
precisely the nine topological classes (\ref{bjoyce}) of
G$_2$ manifolds found by Joyce. 

\subsection{Mirror symmetry for G$_{\mathbf 2}$ manifolds}
\label{g2.mirror}

In this subsection we will show that T-duality on three suitably
chosen coordinates  generates a nontrivial automorphism of one, say
the right-moving, extended chiral algebra of the G$_2$
compactification. In analogy to the \CY case we call this automorphism 
the `mirror automorphism'. The two theories related by this T-duality
are thus physically equivalent. As we shall show, depending on the
specific choice of the coordinates, this transformation either
reverses {\it all} discrete torsion signs, or {\it none}. In the
former case, the corresponding mirror map then relates IIA/IIB string 
theory on $Y_\ell$ to IIB/IIA string theory on $Y_{8-\ell}$; this is
the generalisation of the mirror symmetry mentioned in 
\cite{Acharya1,Acharya2} to $\ell\ne 0$. In the second case, the
mirror map relates IIA/IIB string theory on $Y_\ell$ to IIB/IIA string
theory on the same manifold $Y_\ell$; this is the mirror symmetry
suggested in \cite{PT}. It is very satisfying that both these mirror 
symmetries have an interpretation in terms of the automorphism of the
extended G$_2$ algebra.

The extended chiral algebra of a string moving on a compact G$_2$
manifold 
\cite{Shatashvili:1994zw} consists of an $\cN=1$ superconformal
algebra generated by the stress energy tensor $T$ and the supercurrent
$G$, that is extended by a real current $\Phi$ of conformal weight
$h_{\Phi}=3/2$, a current $X$ of conformal weight $h_X=2$ and their
superpartners $K=\{G,\Phi\}$ and $M=[G,X]$ respectively. 
The operator $\Phi$ corresponds to the 3-form defining the
G$_2$ structure on the target space. 
In our free field representation, the relevant currents look like
\begin{align}
T&= \frac{1}{2}\sum_{j=1}^7 :\partial x_j \partial x_j : 
-\frac{1}{2} \sum_{j=1}^7 : \psi^j \partial \psi^j :\,, \qquad
G= \sum_{j=1}^7 : \psi^j \partial x_j : \,, \non \\
\Phi&=\psi^1\psi^3\psi^6+\psi^1\psi^4\psi^5+\psi^2\psi^3\psi^5
-\psi^2\psi^4\psi^6+\psi^1\psi^2\psi^7+\psi^3\psi^4\psi^7
+\psi^5\psi^6\psi^7 \,, \label{Phi} \\
\intertext{and}
X&= -\psi^2\psi^4\psi^5\psi^7
-\psi^2\psi^3\psi^6\psi^7
-\psi^1\psi^4\psi^6\psi^7
+\psi^1\psi^3\psi^5\psi^7 \non \\
& ~~~~-\psi^3\psi^4\psi^5\psi^6 
-\psi^1\psi^2\psi^5\psi^6 
-\psi^1\psi^2\psi^3\psi^4
-\frac{1}{2} \sum_{j=1}^7 : \psi^j \partial \psi^j : \,,\label{X}
\end{align}
which are all preserved by the orbifold action (\ref{g2orb2}).
The algebra they satisfy has been worked out in
\cite{Shatashvili:1994zw}. Of course there is again an isomorphic
right-moving algebra. 

This extended chiral algebra has two automorphisms. The first is the
fermion parity $\psi^j\mapsto -\psi^j$ under which the operators have the
eigenvalues
\begin{equation}
\begin{array}{c|cccccc}
{} & ~T~ & ~G~ & ~\Phi ~ & ~X~ & ~K~ & ~M~ \\ \hline
\mbox{fermion parity}~ & + & - & - & + & + & - \\
\end{array} ~.
\end{equation}
The other automorphism is more interesting,
\begin{equation}
\begin{array}{c|cccccc}
{} & ~T~ & ~G~ & ~\Phi ~ & ~X~ & ~K~ & ~M~ \\ \hline
\mbox{mirror}_{G_2}~ & + & + & - & + & - & + \\
\end{array} \,. \label{aut.g2}
\end{equation}
It leaves the $\cN=1$ superconformal subalgebra generated by $T$ and
$G$ invariant but reverses the operator $\Phi$ and its superpartner
$K$. It is the natural analogue of the mirror automorphism of
the \CY algebra (\ref{aut2}), and we shall therefore call it the 
{\it mirror automorphism}. 

For one class of compact G$_2$ manifolds, namely the manifolds 
$Y=({\rm CY}_3\times S^1)/\Bz_2$ where the $\Bz_2$ acts as a real
structure on CY$_3$ and as an inversion $x_7\mapsto -x_7$ on the
circle, the G$_2$ mirror automorphism is actually generated by 
the CY mirror automorphism (applied to the CY$_3$ part of the above
space). To see this, one expresses the generators of the extended 
chiral algebra for the G$_2$ compactification in terms of those of the
\CY manifold and the $S^1$-compactification (where the latter are
described by $\partial x^7$ and $\psi^7$) 
\cite{Figueroa-O'Farrill:1997hm},
\begin{align}
T&=T_{\rm CY}+\frac{1}{2}:\partial x_7 \partial x_7 : 
-\frac{1}{2} : \psi^7 \partial \psi^7 :\, , \qquad
G= G_{\rm CY} + :\psi^7 \partial x_7 :\, , \non \\
\Phi &= \im(\Om)+:J \psi^7: \, , \qquad
X= :\re(\Om)\psi^7: + \frac{1}{2} :JJ:
-\frac{1}{2}:\partial \psi^7 \, \psi^7: \, , \non \\
K&= \im(\Psi)+:J\partial x_7:+:G^{\prime}_{\rm CY} \psi^7: \, , \non \\
M&=:\re(\Psi)\psi^7:-:\re(\Om)\partial x_7: +:\partial x_7\partial \psi^7:
+:JG_{\rm CY}^{\prime}:-\frac{1}{2}\partial G_{\rm CY} \, . \non
\end{align}  
It is then easy to see that the application of the automorphism
(\ref{aut2}) to the \CY generators gives rise to the automorphism
(\ref{aut.g2}) on the G$_2$ generators.  

Next let 
\begin{align}
{\cal I}_3^+ &=\{(2,4,6),(2,3,5),(1,2,7)\} \,, \label{Iplus} \\
{\cal I}_3^- &=\{(1,3,6),(1,4,5),(3,4,7),(5,6,7)\}  \label{Iminus}
\end{align}
with $\cI_3=\cI_3^+\cup \cI_3^-$ be the index set appearing in (\ref{Phi}).  
Then since T-duality on the coordinate $x_j$ reverses
the right-moving currents $\overline{\partial} x_j$ and $\tpsi^j$ but
leaves the left-moving currents $\partial x_j$ and $\psi^j$ invariant,
we see that simultaneous T-duality on $x_{j_1},x_{j_2},x_{j_3}$ for
$(j_1,j_2,j_3)\in \cI_3$ generates the mirror automorphism (\ref{aut.g2})
on the right-moving chiral algebra while being the identity on the
left-moving one.\footnote{It is easy to check that, apart from those
index sets appearing in (\ref{Iplus}) and (\ref{Iminus}), the only
other T-duality transformation which has this effect is the T-duality
transformation on all seven coordinates. For the following discussion
it behaves as an element in $\cI_3^+$.}   

Next we want to show that for $(j_1,j_2,j_3)\in \cI_3^+$ the T-dualities
leave all discrete torsion phases invariant, while for 
$(j_1,j_2,j_3)\in \cI_3^-$ the T-dualities reverse all eight discrete
torsion phases. In the former case, this therefore maps the manifold
$Y_\ell$ to itself, 
whereas in the second case it exchanges $Y_\ell$
with $Y_{8-\ell}$ (while in both cases exchanging type IIA and type IIB 
strings).

\begin{figure}[htbp]
\centering
\input{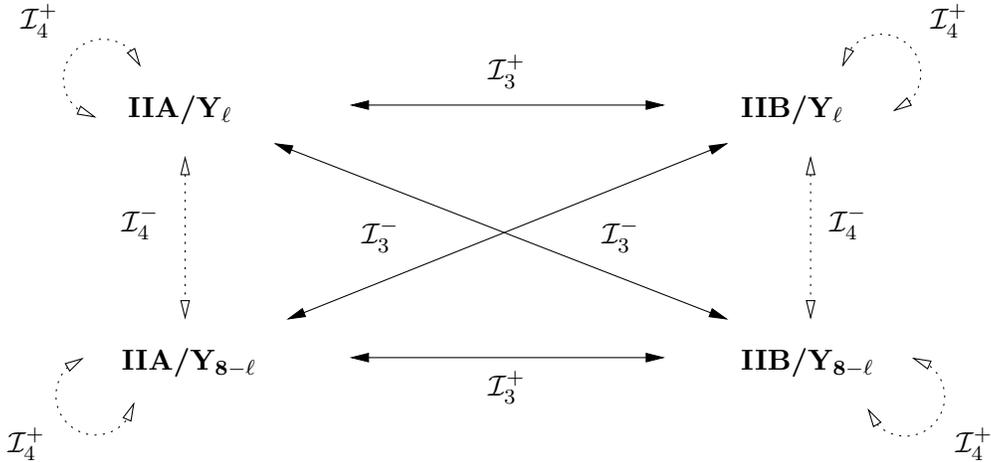}
\caption{Dualities generated by T-duality on the coordinates of
$\cI_{3,4}^{\pm}$.}
\label{dual}
\end{figure}

In order to see this, we need to study how the
action of these T-dualities modifies the action of $\alpha\beta$ 
in the $\gamma$-twisted RR sector. [This is one of the places where
the discrete torsion signs appear; it is not difficult to see that all
other sectors behave accordingly.] For the $\gamma$-twisted 
RR sector we have fermionic zero modes for $i=2,4,6$; of these,
$\alpha\beta$ inverts the directions $i=4,6$. On the RR ground states
in $\cH_{\ga;\tf}$,
it can thus be represented in terms of fermionic zero modes as  
\begin{equation}
\alpha\beta = \frac{1}{4} \psi^4_0 \psi^6_0 \tpsi^4_0 \tpsi^6_0 
\cdot \eps_{\ga;\tf} \,.
\end{equation}
T-duality in the direction $j$ introduces a sign for $\tpsi^j_0$, but
none for $\psi^j_0$. By inspection of (\ref{Iplus}) and (\ref{Iminus})
it then follows that the T-dualities associated to $\cI_{3}^+$ do
not modify the action of $\alpha\beta$, while those associated to
$\cI_{3}^-$ do. Since this analysis applies uniformly to all fixed
points, it follows that in the second case {\it all} discrete torsion
signs are reversed, and thus that the corresponding duality relates
$Y_\ell$ to $Y_{8-\ell}$. These dualities are summarised in Figure
\ref{dual}.

Obviously, we can also combine two (distinct) such transformations; the 
resulting T-duality transformation inverts then precisely four  
coordinates. These T-duality transformations fall naturally into two
classes $\cI_4^\pm$, depending on whether the two T-dualities in
$\cI_3$ lie both in the same set $\cI_3^\pm$, or in different sets, 
\begin{align}
{\cal I}_4^+&=\{(1,3,5,7),(1,4,6,7),(3,4,5,6)\} \, , \label{I4plus} \\
{\cal I}_4^-&=\{(2,4,5,7),(2,3,6,7),(1,2,5,6),
(1,2,3,4) \} \,. \label{I4minus}
\end{align}
All of these transformations obviously leave both chiral algebras
invariant, but the transformations in $\cI_4^-$ invert all discrete
torsion signs, while those in $\cI_4^+$ do not modify them. As a
consequence, these transformations then relate IIA/IIB theory on 
$Y_\ell$ to IIA/IIB theory on $Y_\ell$ (in the case of $\cI_4^+$) or 
on $Y_{8-\ell}$ (in the case of $\cI_4^-$). These dualities (which 
are a direct consequence of the mirror symmetries associated to 
$\cI_3$) are also summarised in Figure \ref{dual}. Some of them were
considered before in \cite{Acharya1,Acharya2}, and they may be
related to the mirror symmetries suggested in \cite{Lee:2002fa}.

\section{Conclusions}

In this paper we have shown how all nine
G$_2$ manifolds of Joyce coming from the resolution of $T^7/\Bz_2^3$ 
can be realised in terms of a string theoretic orbifold
construction. This involved a generalisation of discrete torsion where
the action of $g$ in the $h$-twisted sector is not just modified by an
overall phase, but by phases that are in general different for 
states that are associated to different fixed points (or twist fields)
in the $h$-twisted sector. 
We have shown that the resulting theories
are still modular invariant at one loop 
provided that these generalised discrete 
torsion phases satisfy certain constraints (that we have solved). 
It would be interesting to understand whether (and if so which)
conditions arise from analysing higher loop modular invariance. 
It would also be interesting to see whether there are other instances
where this generalisation of discrete torsion is of
significance. Finally, it would be desirable to understand the
constraints of modular invariance, at least for some classes of
examples, in a more conceptual fashion. 

We have also proposed that, from a conformal field theoretic point of
view, mirror symmetry for G$_2$ manifolds should be understood as a
consequence of the `mirror automorphism' of the extended G$_2$
algebra. This is the natural generalisation of 
how mirror symmetry arises for \CY manifolds \cite{Lerche:1989uy}. For
the example considered in this paper, we have shown that this point of
view precisely reproduces the symmetry proposed in \cite{PT}, as well
as a generalisation of the mirror symmetry found in
\cite{Acharya1,Acharya2}. 

It would be interesting to understand in detail the relation of 
our proposal to the ideas of \cite{Lee:2002fa}, where
a duality between G$_2$ manifolds is conjectured using fibrewise
Fourier-Mukai transformation on (co)associative fibres, and to study the
Yukawa couplings they propose from a string theory point of view.
It would also be interesting to understand the relation to the work of
\cite{GYZ}, that allows to establish dualities between G$_2$ manifolds
by relating M-theory on them to various compactifications of
ten-dimensional string theory. 
A more direct link exists to mirror symmetry for $\cN=1$ supersymmetric
flux compactifications to four dimensions considered in \cite{SLMW}. 
The half-flat manifolds appearing there give rise to G$_2$ manifolds 
upon compactification on an additional $S^1$. 
Mirror symmetry of the flux backgrounds should then induce mirror 
symmetry of these G$_2$ manifolds. Finally, it would be 
interesting to study this mirror symmetry for other G$_2$ manifolds
for which a conformal field theory description is available  
\cite{Roiban:2002iv,ES,BB,RW,ES1,Noyvert,SY,ESY}. 
It may also be interesting to see whether there is a similar 
construction for Spin(8) manifolds.

\vspace*{10mm}
\noindent {\bf Acknowledgements}

\noindent
We would like to thank Hanno Klemm, Wolfgang Lerche, Tako Mattik,
Andreas Recknagel, Daniel Roggenkamp, Fred Roose and 
Daniel Waldram for useful discussions. 
We would also like to thank the referee for drawing our attention to
the problem of modular invariance at higher genus, and 
to Bobby Acharya for helpful discussions on this issue.


\appendix
\section{Discrete torsion in the $\mathbf{T^7/\Bz_2^3}$-orbifold}
\label{app1}

In this appendix we analyse the possible generalised discrete torsion
phases for the example of the orbifold $T^7/\Bz_2^3$. We shall first
analyse, for each twisted sector, how many phases can be introduced
that satisfy (\ref{repprop}), as well as the constraint that arises
from the modular $T$-transformation, namely $\eps_f(g,g)=1$. Once
the possible phases have been determined, we shall then consider the
constraints that arise from the modular $S$-transformation. 
\smallskip

In the {\bf $\al$-twisted sector} the highest weight states  
are labelled by $|m,n;f\rangle_{\al}$ where $m,n$ are the
integral momentum/winding modes in  the $\al$-untwisted directions
$x_{5,6,7}$ and $f=1,\ldots,16$ label the different twist fields of
lowest conformal dimension.
We shall only consider the generalisation of discrete torsion where
the phases depend on $f$, but not on any other parameters (such as the 
winding or momentum modes). Thus we may restrict ourselves to
considering the ground state with $(m,n)=(0,0)$, leaving us with a
16-dimensional space $|0,0;f\rangle_{\al}$. The 16 twist fields are in
one-to-one correspondence with the fixed points $x_j\in\{0,1/2\}$ of
the action $x_j\mapsto -x_j$ of $\al$ on the first four coordinates
$x_{1,2,3,4}$.  The action of $\G$ on the label $f$ can thus be
inferred from how $\G$ permutes these 16 fixed points: it groups them
into four orbits labelled by the fixed points in the coordinates
$x_{3,4}$, each orbit consisting of the four fixed points in the
coordinates $x_{1,2}$. The irreducible  representations on the twist
fields are therefore 4-dimensional. Labelling them by the
$(x_1,x_2)$-coordinates of the associated fixed points as
\begin{align}
& |1\rangle =\left(x_1=0,x_2=0\right)\,, \quad 
& |2\rangle = \left(x_1=0,x_2=\frac{1}{2}\right)\,, \\
& |3\rangle =\left(x_1=\frac{1}{2},x_2=0\right)\,, \quad 
& |4\rangle = \left(x_1=\frac{1}{2},x_2=\frac{1}{2}\right)\,,
\end{align}
the generators of $\G$ act as follows,
\[
\al=\ident \,, \quad \beta=\left(\begin{array}{c@{\leftrightarrow}c}
|1\rangle & |2\rangle \\
|3\rangle & |4\rangle \\
\end{array}\right)\,, \quad
\ga=\left(\begin{array}{c@{\leftrightarrow}c}
|1\rangle & |3\rangle \\
|2\rangle & |4\rangle \\
\end{array}\right) \,.
\]
Let 
\[
H:= \begin{pmatrix} 0 & 1 \\ 1 & 0 \end{pmatrix} \,,
\]
then by possibly redefining the basis vectors by phases if necessary,
we can always set 
\[
\beta=\left(\begin{array}{c|c}
H & 0 \\ \hline
0 & H \\
\end{array}\right) 
\quad \mbox{and}\quad
\ga=\left(\begin{array}{cc|cc}
0 & 0 & 1 & 0 \\
0 & 0 & 0 & e^{i\vth} \\ \hline
1 & 0 & 0 & 0 \\
0 & e^{-i\vth} & 0 & 0 \\
\end{array}\right)\,. 
\]
The constraint $\beta\ga=\ga\beta$ implies $e^{i\vth}=1$, so that
\[
\ga=\left(\begin{array}{c|c} 0 & \ident \\ \hline
\ident & 0 \\ 
\end{array}\right) . 
\]  
Hence we are able to absorb all possible phases by redefining the
basis vectors. Moreover, all elements $g\in\G$ with $g\ne e,\al$ act
non-diagonally on the highest weight vectors, and thus their traces
vanish. 
\smallskip

The {\bf $\beta$-twisted sector} is completely analogous to the
$\alpha$-twisted sector, except that the fixed points in the second
coordinate lie now at $x_2\in\{1/4,3/4\}$. Upon exchanging the roles
of $\al$ and $\beta$ with respect to the $\al$-twisted sector we
obtain the same conclusions as above.
\smallskip

In the {\bf $\ga$-twisted sector} the 16 dimensional space spanned by 
$|0,0;f\rangle_{\ga}$ decomposes into 8 irreducible $\G$-modules
each of dimension 2 and spanned by the fixed points in the
$x_1$-plane, 
\[
|1\rangle = \left(x_1=\frac{1}{4}\right)\,, \quad
|2\rangle = \left(x_1=\frac{3}{4}\right) \,.
\]
The orbifold generators act on them as 
\[
\al=(|1\rangle \leftrightarrow |2\rangle) \,, \quad
\beta=(|1\rangle \leftrightarrow |2\rangle) \,, \quad
\ga=\ident \,.
\]
By a suitable choice of basis we can always set
\begin{equation}
\al=H \quad \mbox{and}\quad
\beta= \begin{pmatrix} 0 & e^{i\phi} \\ e^{-i\phi} & 0 \end{pmatrix}
\,. 
\end{equation}
The constraint $\al\beta=\beta\al$ then imposes $\phi\in\{0,\pi\}$, so
that 
\begin{equation}
\beta=\eps_{\ga} H \quad \mbox{with}\quad \eps_{\ga}=\pm 1 .
\end{equation}
Thus we have one sign degree of freedom $\eps_{\ga;\tf}=\pm 1$
for each of the eight irreducible representations labelled by
$\tf=1,\ldots,8$.  

Finally, apart from $e,\ga$ there are two further group elements that
act diagonally on the 16-dimensional space spanned by the
$|0,0;f\rangle_{\ga}$. These are $\al\beta$ and $\al\beta\ga$, and their
traces equal
\begin{equation}
\Tr_{\{|0,0;f\rangle_{\ga}\}} (\al\beta )
=\Tr_{\{|0,0;f\rangle_{\ga}\}} (\al\beta\ga) =
2\sum_{\tf=1}^8\eps_{\tf}^{(\ga)} \,.
\end{equation}
\smallskip  

Next consider the {\bf $\al\beta$-twisted sector}. In this (as well as
all the following sectors) there is always one winding mode that takes
values in $\Bz+\frac{1}{2}$; for the case of the $\al\beta$-twisted
sector this is the winding mode $n_2$. Since the orbifold generators
invert this winding number, the ground states are now parametrised by 
$n_2=\pm \frac{1}{2}$. Thus we need to look at a 32-dimensional space
spanned by $|n_2=1/2;f\rangle_{\al\beta}$ and 
$|n_2=-1/2;f\rangle_{\al\beta}$, where $f=1,\ldots,16$ labels the 
16 twist fields corresponding to the fixed points of $x_j\mapsto -x_j$
for $j=3,4,5,6$. 
This space splits into 16 two-dimensional irreducible
representations spanned by the two values for $n_2$, 
\[
|1\rangle = \left(n_2=\frac{1}{2}\right)\,, \quad
|2\rangle = \left(n_2=-\frac{1}{2}\right)\,,
\]
on which the orbifold generators act as 
\[
\al=(|1\rangle \leftrightarrow |2\rangle) \,, \quad
\beta=(|1\rangle \leftrightarrow |2\rangle) \,, \quad
\ga=\mbox{diagonal} \,.
\]
By a suitable choice of basis we can set
\[
\al=H=\beta \quad (\mbox{so that}~\al\beta=\ident) \quad \mbox{and}
\quad \ga=\begin{pmatrix} \eps_1 & 0 \\ 0 & \eps_2 \end{pmatrix} \,, 
\]
with $\eps_i=\pm 1$. The vanishing of the commutator of $\ga$ with $\al$
or $\beta$ requires $\eps_1=\eps_2$, and therefore
\[
\ga=\eps_{\al\beta} \ident \,.
\]
Incidentally, the same condition is also needed in order for the
discrete torsion phases to be independent of the winding mode $n_2$.
Each of the 16 irreducible representations has one sign, and thus we
have in total
\begin{equation}
\mbox{16 signs:}\qquad\qquad   
\eps_{\al\beta;\tf}=\pm 1 \,, \quad
\tf=1,\ldots,16 \,.
\end{equation}

Apart from $e,\al\beta$ there are two further group elements
that act diagonally on the 32-dimensional space spanned by 
$|n_2=1/2;f\rangle_{\al\beta}$ and
$|n_2=-1/2;f\rangle_{\al\beta}$. These are $\ga$ and $\al\beta\ga$,
and their traces over the subspace of fixed $n_2$ are 
\begin{equation}
\Tr_{\{|n_2=1/2;f\rangle_{\al\beta}\}} (\ga)
=\Tr_{\{|n_2=1/2;f\rangle_{\al\beta}\}} (\al\beta\ga)
=\sum_{\tf=1}^{16}\eps_{\tf}^{(\al\beta)} \,.
\end{equation}
The result for $n_2=-1/2$ is obviously the same.
\smallskip

In the {\bf $\al\ga$-twisted sector} the ground states are spanned by
the 16 twist fields corresponding to the fixed points in the
$x_{2,4,5,7}$-plane, and the minimal winding numbers $n_1=\pm 1/2$
along the first direction. This space splits into 8 irreducible
representations of dimension 4 each, that are spanned by the fixed
points in the $x_2$-plane and the winding numbers $n_1=\pm 1/2$,
\begin{align}
&|1\rangle = \left(x_2=0,n_1=-\frac{1}{2}\right)\,,\quad
&|2\rangle = \left(x_2=0,n_1=\frac{1}{2}\right) \,,\non \\
&|3\rangle = \left(x_2=\frac{1}{2},n_1=-\frac{1}{2}\right)\,, \quad
&|4\rangle = \left(x_2=\frac{1}{2},n_1=\frac{1}{2}\right) \,.\non
\end{align}
The orbifold generators act on these states as 
\[
\al=\left(\begin{array}{c@{\leftrightarrow}c}
|1\rangle & |2\rangle \\
|3\rangle & |4\rangle \\
\end{array}\right) \,, \quad
\beta=\left(\begin{array}{c@{\leftrightarrow}c}
|1\rangle & |4\rangle \\
|2\rangle & |3\rangle \\
\end{array}\right) \,, \quad 
\ga=\left(\begin{array}{c@{\leftrightarrow}c}
|1\rangle & |2\rangle \\
|3\rangle & |4\rangle \\
\end{array}\right) \,.
\]
By a suitable choice of basis we can set
\begin{equation}
\al= \begin{pmatrix} H & 0 \\ 0 & H \end{pmatrix}=\ga
\quad (\mbox{so that }\quad \al\ga=\ident) \,, \qquad \mbox{and}\quad
\beta= \begin{pmatrix} 0 & 0 & 0 & 1 \\
 0 & 0 & e^{i\vth} & 0 \\
 0 & e^{-i\vth} & 0 & 0 \\
 1 & 0 & 0 & 0 \end{pmatrix} \,.
\end{equation}
The constraint $\al\beta=\beta\al$ then imposes 
$e^{i\vth}=1$, so that
\begin{equation}
\beta=\begin{pmatrix} 0 & H \\ H & 0 \end{pmatrix} \,.
\end{equation}
Hence we are again able to absorb all these phases into a redefinition
of the basis vectors. Moreover, all elements $g\in\G$, except for
$g=e$ and  $g=\al\ga$ act non-diagonally on these highest weight
vectors, and thus their traces vanish.
\smallskip

The {\bf $\beta\ga$-twisted sector} is completely analogous to the 
$\al\ga$-twisted sector, except that the fixed points in the second
coordinate now lie at $x_2\in\{1/4,3/4\}$. Upon exchanging the roles
of $\al$ and $\beta$ with respect to the $\al\ga$-twisted sector we
obtain the same conclusion.
\smallskip

Lastly we consider the {\bf $\al\beta\ga$-twisted sector}. The ground
states are again spanned by the 16 fixed points in the
$x_{1,4,6,7}$-plane, and the minimal winding numbers $n_2=\pm 1/2$ in
the second direction. It splits into 8 irreducible representations of
dimension 4 each, that are spanned by the fixed points in the
$x_1$-plane and the winding number $n_2=\pm 1/2$,
\begin{align}
&|1\rangle = \left(x_1=\frac{1}{4},n_2=-\frac{1}{2}\right)\,,\quad
&|2\rangle = \left(x_1=\frac{1}{4},n_2=\frac{1}{2}\right) \,,\non \\
&|3\rangle = \left(x_1=\frac{3}{4},n_2=-\frac{1}{2}\right) \,, \quad
&|4\rangle = \left(x_1=\frac{3}{4},n_2=\frac{1}{2}\right) \,.\non
\end{align}
The orbifold generators act on them as 
\[
\al=\left(\begin{array}{c@{\leftrightarrow}c}
|1\rangle & |4\rangle \\
|2\rangle & |3\rangle \\
\end{array}\right) \,, \quad
\beta=\left(\begin{array}{c@{\leftrightarrow}c}
|1\rangle & |4\rangle \\
|2\rangle & |3\rangle \\
\end{array}\right) \,, \quad 
\ga=\mbox{diagonal} \,.
\]
By a suitable choice of basis we can set
\begin{equation}
\al= \begin{pmatrix} 0 & 0 & 0 & 1 \\
0 & 0 & 1 & 0 \\
0 & 1 & 0 & 0 \\
1 & 0 & 0 & 0 \end{pmatrix}  \,, \;\;\;
\beta= \begin{pmatrix} 0 & 0 & 0 & e^{i\phi} \\ 
0 & 0 & e^{i\vth} & 0 \\
0 & e^{-i\vth} & 0 & 0 \\
e^{-i\phi} & 0 & 0 & 0 \end{pmatrix} \,,
\quad \mbox{and} \;\;\;
\ga= \begin{pmatrix} \tilde{\eps}_1 & 0 & 0 & 0 \\ 
0 & \tilde{\eps}_1 & 0 & 0 \\
0 & 0 & \tilde{\eps}_2 & 0 \\
0 & 0 & 0 & \tilde{\eps}_2 \end{pmatrix} \,,
\end{equation}
where we have used that the action of $\gamma$ should be independent
of $n_2$. The constraint $\al\beta=\beta\al$ then imposes 
$e^{i\phi}=e^{-i\phi}=\pm 1=: \eps_1$ and
$e^{i\vth}=e^{-i\vth}=\pm 1=: \eps_2$, whereas
the constraint $\al\ga=\ga\al$ imposes
$\tilde{\eps}_1=\tilde{\eps}_2=:\eps_{3}$. The $T$-constraint
$\ident\stackrel{!}{=}\al\beta\ga$
then demands $\eps_1=\eps_{3}$ and $\eps_2=\eps_3$, so that
\begin{equation}
\beta= \begin{pmatrix} 0 & 0 & 0 & \eps_1 \\ 
0 & 0 & \eps_1 & 0 \\
0 & \eps_1 & 0 & 0 \\
\eps_1 & 0 & 0 & 0 \end{pmatrix}\quad \mbox{and}\quad
\ga= \begin{pmatrix} \eps_1 & 0 & 0 & 0 \\ 
0 & \eps_1 & 0 & 0 \\
0 & 0 & \eps_1 & 0 \\
0 & 0 & 0 & \eps_1 \end{pmatrix}\,. 
\end{equation}
Each irreducible representation therefore has one sign degree of
freedom $\eps_{\al\beta\ga}=\pm 1$. On the 32-dimensional 
space of ground states
we thus have the freedom to choose
\begin{equation}
\mbox{8 signs:} \qquad\qquad
\eps_{\al\beta\ga;\tf}=\pm 1, \quad \tf=1,\ldots,8 \,.
\end{equation} 
Finally, apart from $e,\al\beta\ga$ there are two further group
elements that act diagonally on the 32-dimensional space of ground
states. These are $\ga$ and $\al\beta$, and their traces over the
subspace of fixed $n_2$ equals
\begin{equation}
\Tr_{\{|n_2=1/2;f\rangle_{\al\beta\ga}\}}(\ga )
=\Tr_{\{|n_2=1/2;f\rangle_{\al\beta\ga}\}}(\al\beta)
=2\sum_{\tf=1}^{8}\eps_{\al\beta\ga;\tf}\,.
\end{equation}
Again, the results for $n_2=-1/2$ are obviously the same.
\smallskip

In summary, we therefore have 8 signs 
$\eps_{\ga;\tf}=\pm 1$ for $\tf=1,\ldots,8$, in the $\ga$-twisted
sector;  
another 8 signs $\eps_{\al\beta\ga;\tf}=\pm 1$ for $\tf=1,\ldots,8$, 
in the $\al\beta\ga$-twisted sector;  and 
16 signs 
$\eps_{\al\beta;\tf}=\pm 1$ for $\tf=1,\ldots,16$,
in the $\al\beta$-twisted sector. We are now in a position to analyse
the constraint that arises from the modular $S$-transformation. The
non-trivial conditions come from
\begin{align}
&\begin{picture}(10,10)(0,0)
\put(5,0){\framebox(4,4){}}
\put(6,-3){\mbox{${}_{\ga}$}}
\put(0,2){\mbox{${}_{\al\beta}$}}
\end{picture}(q(-1/\tau))=
\begin{picture}(10,10)(0,0)
\put(5,0){\framebox(4,4){}}
\put(5,-3){\mbox{${}_{\al\beta}$}}
\put(2,2){\mbox{${}_{\ga}$}}
\end{picture}(q(\tau)) \,, 
\qquad
\begin{picture}(10,10)(0,0)
\put(5,0){\framebox(4,4){}}
\put(6,-3){\mbox{${}_{\ga}$}}
\put(-1,2){\mbox{${}_{\al\beta\ga}$}}
\end{picture}(q(-1/\tau))=
\begin{picture}(10,10)(0,0)
\put(5,0){\framebox(4,4){}}
\put(5,-3){\mbox{${}_{\al\beta\ga}$}}
\put(2,2){\mbox{${}_{\ga}$}}
\end{picture}(q(\tau)) \,, \non \\
&\begin{picture}(10,10)(0,0)
\put(5,0){\framebox(4,4){}}
\put(5,-3){\mbox{${}_{\al\beta\ga}$}}
\put(0,2){\mbox{${}_{\al\beta}$}}
\end{picture}(q(-1/\tau))=~
\begin{picture}(10,10)(0,0)
\put(5,0){\framebox(4,4){}}
\put(6,-3){\mbox{${}_{\al\beta}$}}
\put(-1,2){\mbox{${}_{\al\beta\ga}$}}
\end{picture}(q(\tau)) \,, 
\end{align}
and they lead to the constraints
\begin{equation}
2\sum_{\tf=1}^8 \eps_{\ga;\tf}
=\sum_{\tf=1}^{16}\eps_{\al\beta;\tf}
=2\sum_{\tf=1}^8 \eps_{\al\beta\ga;\tf}\,, \label{Sg2b}
\end{equation}
as claimed in the main part of the paper.



\end{document}